\documentstyle[aaspp4,natbib209]{article}
\def\eg{{\it e.g.,~}}
\def\ie{{\it i.e.,~}}

\def\eg{{\it e.g.,~}}
\def\ie{{\it i.e.,~}}

\def\etal{{\it et al.~}}

\begin{document}
\title{ A PLASMA PRISM MODEL FOR AN ANOMALOUS DISPERSION EVENT IN 
THE CRAB PULSAR}

\author{ D. C. Backer, T. Wong, \& J. Valanju }
\medskip
\affil{ Astronomy Department \& Radio Astronomy Laboratory,
University of California, Berkeley, CA \\
email: dbacker@astro.berkeley.edu
}

\begin{abstract}
In 1997 October daily monitoring observations of the Crab pulsar
at 327 MHz and 610 MHz with an 85ft telescope in Green Bank, WV showed
a jump in the dispersion measure by 0.1 cm$^{-3}$ pc. Pulses were seen
simultaneously at both old and new dispersions over the course of several
days. During the dispersion jump the pulsed flux diminished by an order
of magnitude.  In the months before this event the average pulse profiles
contained faint ``ghost'' pulse components offset in phase from the regular
main pulse and interpulse components by a nearly frequency independent
time delay that quadratically diminished to zero over a month. After the dispersion
event there was an order of magnitude increase in the level of scattering,
as measured by pulse broadening at 327 MHz. There was also a
curious shift in the rotational phase, a slowing down, at both frequencies
at the time of the dispersion jump which we associate with intrinsic
timing noise. All of the observed phenomena except this slowing down
can be explained by the variable perturbing 
optics of a triangular plasma prism that is located in the filamentary
interface between the synchrotron nebula and the supernova ejecta
and which crosses the line of sight over a period of months.
The required density, scale length and velocity are reasonable
given previous observations and analysis of these filaments.
Our study thus provides a probe of the plasma column on scales of 
30 microarcsecond to 3 milliarcsecond ($10^{12-14}$ cm)
which complements scales accessible to optical emission line studies 
with HST resolution ($10^{16-18}$ cm). In combination both observations
provide a detailed look at a sample
of the interface region that can be matched
statistically to the results of numerical simulations. 
\end{abstract}


\section{INTRODUCTION}

Observations of pulsar signals at multiple frequencies allow measurements
of a number of properties of the intervening plasma. The 
quadratic dependence of arrival times  on radio frequency yields the column 
density of electrons, or
dispersion measure, along the path. At low frequencies pulses are
broadened by multipath propagation (diffraction) effects owing to the
presence of phase perturbations 
on transverse scales smaller than the Fresnel radius. These perturbations
are the result of microscale electron density fluctuations.
When the perturbations are confined to a small
region along the sight line and have a wide extent transverse to the
path, the broadening function is exponential. Phase pertubations
that are large with respect to the Fresnel radius lead to refractive
changes both in the path direction and in the signal amplitude. 
Most intervening media are birefringent, and as a result
the reference angle of linear polarization undergoes
Faraday rotation as a function of radio frequency.
Temporal changes in all of these quantities result
from the net motion of the sight line relative to the perturbing plasma.
Amplitude changes in time or in frequency
are called {\it scintillation}, while all diffractive
and refractive phenomena are the result of {\it scattering} of electromagnetic
waves by electron density structures; scintillation and scattering are
not synonyms. The temporal changes provide
a sensitive probe of the smallest known structures in the intervening
medium which have eluded physical description (\eg recent studies by
\citeauthor{Sridhar94} \citeyear{Sridhar94},
\citeauthor{Goldreich97} \citeyear{Goldreich95}, \citeyear{Goldreich97})
since the beginning of pulsar research in 1968. The turbulent energy
in these small scale structures in the interstellar medium may be
a significant source of heat (\eg \citeauthor{Minter97a} \citeyear{Minter97a,Minter97b})
as well as source of scattering of cosmic rays ({\eg \citeauthor{Jokipii88}
\citeyear{Jokipii88}).
One difficulty in the understanding of the small scale strutures in
the interstellar medium is the uncertainty about the location of the
structures along the line of sight 
and therefore an uncertainty in the mean properties of the perturbed plasma.
In this paper we describe the recent variable propagation effects in the Crab 
pulsar and argue, as others have previously, that they result from plasma 
associated with nebula. In this case we 
 have significant other information about the state of the material from
optical observations which can aide theoretical analysis.

The propagation variations we will be discusssing have a historical precedence.
In 1974-75 the dispersion measure, rotation measure and scattering of the
Crab pulsar displayed extreme activity \cite[]{Lyne75,Isaacman77,Rankin88}. 
The dispersion measure rose by 0.07 cm$^{-3}$pc and the Faraday rotation
rose by 2 radians m$^{-2}$ over several months. The scattering increased
by an order of magnitude, and there was a deep, presumably refractive,
amplitude null.
These disturbances in the observed emission from the pulsar were
ascribed to propagation through perturbed thermal plasma associated with the 
Crab nebula. 
\cite{Rankin73} had already argued 
on the basis of a two scattering screen model
that the smaller variations of
dispersion and Faraday rotation which they had observed
were the result of nebular material. The variable screen
was associated with the nebula, while the constant screen was associated 
with the general interstellar 
medium. \cite{Vandenberg76} compared VLBI observations of the apparent
size to the scattering time scale and also showed the need for a 
scattering screen near the nebula. These authors suggested that the
observed effects were the result of propagation through fine structure
in the well-known filamentary material which was known to have
structure on many length scales.
More recently, precision observations of other pulsars have determined
the level of dispersion measure variations in the general interstellar
medium and found them to be significantly below that seen in the Crab
\cite[]{Phillips91,Phillips92,Backer93}.

In recent years there has been a second episode of large variations 
in the dispersion measure and scattering time scale 
\cite[]{Backer96}. Column density
variations of 0.1 cm$^{-3}$ pc over time scales of months are seen.
Starting farthest from the pulsar, 
potential sites for the perturbing plasma (and their radii from the pulsar)
are:
(a) the supernova blast wave forward and reverse shocks ($\sim$10 pc); 
(b) the shock and Rayleigh-Taylor unstable shell and shell
fragments formed by the expanding synchrotron nebula (1-1.5 pc); 
(c) the synchrotron emitting particles (0-1 pc);
and (d) the immediate environs of pulsar and relativistic pulsar
wind shock (0.2 pc). Our estimate of the
density of the synchrotron emitting electrons is $10^{-5}$ cm$^{-3}$, and
so even the electrons in the weakly relativistic end of their energy distribution 
produce negligible dispersion in the hot interior of the nebula. 

We will interpret our observations as the result of propagation through
category (b) material,
the unstable shell which forms  filamentary structure, although admittedly this 
is in part because it is the structure about which we know the most from optical 
spectroscopy. Here we will distinguish
between the diffuse high ionization material characterized by
`billowing' structures within a thick envelope region around the synchrotron
nebula \cite[]{Lawrence95,Sankrit98}, and the low ionization, denser
material which has a much smaller filling factor and is found over 
a range of radii.  Both are loosely referred to as `filaments' 
in the literature. \cite{Sankrit98} argue that the
high ionization material is the result of a shock between the synchrotron
emitting internal gas and the external supernova ejecta and conclude that
its radial cross section is $\sim 2\times 10^{15}$ cm.
Rayleigh-Taylor instabilities in the shock interface are the likely
source of the cooler filaments \cite[]{Chevalier75,Hester96}. 
The typical density of this material is
1000 cm$^{-3}$ and the typical scale is $10^{16}-10^{17}$ cm
\cite[]{Fesen92,Lawrence95,Hester96}.  Small structures or small
variations within larger structures found in these optical studies
could easily produce the changes in the 1974
and current episodes of large variability in propagation parameters.
If the variations that we will discuss are the result of changes in the sight 
line column density through this filamentary material,
then the transverse velocity of the
pulsar-Earth sight line relative to the plasma is at least that of the
pulsar, $\sim$135 km s$^{-1}$ \cite[]{Wycoff77,Caraveo99}. The filament motions
are radial \cite[]{Trimble68} to within 100 km s$^{-1}$,
and the shock velocity in the high ionization material is estimated
to be $\sim 150$ km s$^{-1}$ \cite[]{Sankrit98}. We will assume a value of 
200 km s$^{-1}$ for the characteristic transverse velocity of the sight line 
relative to the filaments for conversion from the time domain to the 
spatial domain. In this case, the estimated electron density of the observed
perturbations is $\sim 1500$ cm$^{-3}$, and their transverse length is
$\sim 2\times 10^{14}$ cm for single symmetric structures. 
This density is larger than, but comparable to, the estimates from optical line 
measurements.  Smaller density variations on smaller length scales would
be expected.

In 1997 October amidst the recent episode of large variations in dispersion 
measure and other plasma propagation parameters, a dispersion measure ``jump''
(a sudden change in less than one week)
of 0.12 cm$^{-3}$pc was noticed in measurements with monitoring telescopes
at both the NRAO\footnote{The 
National Radio Astronomy Observatory is a facility of the National Science 
Foundation operated under cooperative agreement by Associated Universities, Inc.}
Green Bank and the University of Manchester Jodrell Bank 
sites \cite[]{Backer00, Smith00}. 
In fact, at the time of the jump pulses were simultaneously detected at both
old and new DMs. Subsequently \citeauthor{Smith00} noticed that prior to the
dispersion event a faint ``ghost'' replica of the pulsed emission following
the main pulse and interpulse components was seen for an interval of about two 
months. The phase of the
ghost emission slowly converged toward that of the regular emission a few weeks
before the dispersion jump. \citeauthor{Smith00} discuss a geomtrical optics
model for variable reflection of the signal from a plasma wall in the nebula, and 
infer a location of 2 pc from the pulsar. 
In this paper we interpret the observations with a full physical optics model
of a plasma structure which has the rough form of an overdensity triangular prism.

While many of the observed phenomena can be explained by a
plasma prism passing through the line of sight at the distance of the Crab
nebula filaments,
the occurrence of an unusual slowing down of the neutron star as well as the small
but more conventional spinup glitch of the star two months later
could provide an argument for consideration of plasma propagation in the
vicinity of the star to allow for a causal connection. The fact that the
ghost components of the main pulse and the interpulse
are not always similar adds further confusion to the filament propagation
interpretation. However, we argue in this paper 
that the propagation and spin disturbances are not causally connected.

In \S 2 we describe the Pulsar Monitoring Telescope facility at NRAO
Green Bank along with our data collection and analysis procedures. In \S 3 we
present a general description of the late 1997 events in the Crab data based
on our observations at 327 MHz and 610 MHz. In a future work we will
present a longer history of propagation parameters and their relation
to the filamentary nebular material.
This is followed in \S
4 by a discussion of the rotation history of the pulsar and in \S 5
by a presentation of our 
plasma prism model that explains most of the features of the event.
In \S 6 we discuss the properties of the ghost emission 
along with our interpretation. In the concluding section we summarize 
our arguments for associating the observed phenomena with propagation effects
in the Crab nebula's filamentary zone,
and pose several questions that would motivate observations
at the time of a future event. 
 
\section{OBSERVATIONS WITH AN 85ft PULSAR MONITORING TELESCOPE}

The observations described below were obtained with an 85ft (26m) diamter
Pulsar Monitoring Telescope in Green Bank, WV. Pulsar observations were
initiated on this telescope by Stinebring, Kaspi and their colleagues
(e.g., \citeauthor{Kaspi92} \citeyear{Kaspi92})
in collaboration with the US Naval Observatory which supported the telescope 
operations
at that time for geodetic work. Since then the Naval Research Laboratory
contributed to
operations, and currently a mixture of resources are applied to keep
the system running.  Room temperature receivers with linearly
polarized feeds at 327 MHz and 610 MHz are mounted off axis and are
then continuously available for 
use by offset pointing. The available bandwidths of the receivers
are 10 MHz and 40 MHz, respectively, while our Crab observations used
8 MHz and 16 MHz, respectively. The Crab pulsar is observed twice
per day at 610 MHz with 10-minute integrations (LST 06:05-06:50 \& 10:35-11:20)
and once per day at 327 MHz with 8.5-minute integrations (LST 06:55-07:35).

Average pulse profiles are obtained from each integration
with the Green Bank-Berkeley Pulsar Processor (GBPP; 
{\it e.g.}, \citeauthor{Backer97} \citeyear{Backer97}). The GBPP coherently 
removes dispersion
from pulsar radiation in real time by convolution in the time domain.
Two othogonally polarized
signals are each separated into 32 frequency channels which span the bandwidths
defined above using analog and digital filter techniques.
The dispersion removal restoring function for each channel
has a maximum length of 1024 samples and the
convolution processor, which employs
a full-custom VLSI chip \cite[]{Kapadia93}, uses low bit quantization. 
The multichannel data from this processor are dedispersed for the time delays
between channels, linearized to remove the effects of quantization, 
normalized by the system temperature and summed over the
two polarizations.
Further reduction includes estimation
of flux, pulse broadening, arrival times and other properties.
Archival data from this monitoring program for the Crab and other pulsars
are available upon request,\footnote{See also http://astro.berkeley.edu/\~{}mpulsar}
and please contact D. Backer if you are interested in use of the
85ft Pulsar Monitoring Telescope for new projects.

\section{ANATOMY OF 327-MHz AND 610-MHz DATA}

In Figure 1 we present a record of the pulse shape at 327 MHz during
the interval of 1997 day 225 to day 365 (1997 Aug 12-Dec 31; MJD 50661-50813). 
The timing model 
that was used to align the daily average pulse profiles is
given in Table 1. The morphology of the Crab's pulsed emission consists of three
components: the precursor (P), main pulse (MP) and interpulse (IP) 
\cite[]{Rankin70}. The properties of these components -- widths,
relative amplitudes and relative locations -- are typically obscured
by the combined effects
of interstellar scattering and instrumental impulse response. 
Our signal processor
minimizes instrumental effects, and we have developed software to deconvolve
the effects of interstellar scattering. Table 2 lists the intrinsic
properties of the pulse components at our two observing frequencies.
In Figure 1 the precursor, main pulse and interpulse components
are overexposed to show a faint, ``ghost'', of the MP
component that appears near phase 0.35 at 1997 day 240 
and drifts toward phase 0.25 at day 275. A similar ghost of the IP component
is evident with similar delays. We don't detect a ghost version of the precursor
component, but it has a lower peak flux and where the MP is strongest, the
location of the expected precursor ghost merges with the regular non-ghost 
emission. The dominant feature of Figure 1 is a jump
in phase and change in signal character around day 300.

Figure 2 summarizes parameters of the 327-MHz profiles during the
interval shown in Figure 1.  We will refer to the signals before
and after the jump as ``old'' (solid line) and ``new'' (dotted line), 
respectively. The old pulse peak amplitude begins steady, has a broad
maximum around day 270 with two local maxima centered
on day 265 and day 275, and then fades away by day 300 (Fig. 2a). 

The new pulse amplitude rises steadily from day 290 to day 365. 
Estimates of the ghost component amplitude are difficult owing to its weakness
and variable structure. The ratio of integrated flux, or pulse energy, of
the ghost emission to the regular emission for the IP ranges from  2-6\%.
The relative arrival times of the
old and new pulses are shown in Figure 2b. This indicates relative
stability apart from the offset between old and new which is the result of
the DM jump.  Note the overlap of the old and the new pulse emission
in Figure 2a,b. This will be interpreted in terms of
{\it two} simultaneous propagation paths during the dispersion jump
in \S 5. As the old pulse amplitude decreases during days 275-300
its position appears at later and later phases with a total
shift between days 295 and 300 of 0.3 ms (Fig. 2b; also Fig. 5 which
will be presented later). We discuss this shift in
terms of an unusually large and rapid excursion of the intrinsic timing
noise in the star in the following section
after presentation of the 610-MHz data. 
The ghost component positions are shown in the second panel (open squares).
These were estimated from the intensity peaks in Figure 1.
Both old and new pulses show evidence of scattering
with broadening by an exponential function whose time constant is given
in Figure 2c. The new pulse is severely scattered in comparison
with that of the old. Scattering by 0.3 ms is typical for this pulsar.
There is little evidence for exponential broadening
in the MP and IP ghost components. 

In Figure 3 we present the corresponding record of the pulse shape at 610 MHz 
during the interval of 1997 day 225 to 1997 day 365. The timing model is the
same as that used for Figure 1. The MP and IP are overexposed to
show the faint precursor near phase 0.22 at 1997 day 225 and 
0.3 at day 365. Intrinsic pulse morphology at 610 MHz is summarized
in Table 2.

Figure 4a shows that the 
amplitudes of the old pulse components rose sharply 
starting on day 270. This rise was followed by local maxima on days 280 and
285 similar to what was reported above in the 327-MHz data, 
but shifted later in time.
After day 290 the old amplitude subsides rapidly out to day 300
where detection is no longer possible. 
There is no evidence of any increase in the 
pulse broadening of this old emission as its amplitude subsides.
The new emission appears diffusely around day 275, and we are
able to fit for amplitudes and scattering times starting on day 315.
The new emission amplitude rises steadily until day 355.

The old emission shows no evidence of scattering, and none would be
expected if we scale the scattering detected at 327 MHz, typically
225 $\mu$s, to 610 MHz using a $\nu^{-4}$ law. The new emission is
definitely scattered. 
From pulse deconvolutions we find exponential time constants ranging 
from 325 to 450 $\mu$s; Figure 4c plots a constant value of 340 $\mu$s. 
This scattering is much {\it larger} than that
based on the 327-MHz results discussed above and using 
a $\nu^{-4}$ law. We return to this discrepancy in \S 5.3.

Starting on day 240 the ``ghost'' emission components at 610 MHz 
follow the MP and IP components and have a typical spread in rotational phase 
of 0.05. The leading
edges of these components, which are their most clearly delineated feature, 
move steadily toward the main pulse and interpulse components (squares
in Fig. 4b). Starting on day 260 the ghost emission components
typically have the shape of an exponentially scattered pulse with time 
scale of approximately 1.2 ms. Figure 4 provides estimates of these
exponential times along with the position and amplitude of the unscattered
ghost emission. These parameters were obtained 
by first removing the strong, weakly scattered
old emission in the MP and IP components,
and then fitting for scattered MP and IP ghost components.
Note that the total pulse energy in the ghost components is typically
half of that in the regular components. This ratio is an order of magnitude
larger than the ratio reported above for 327 MHz.
They appear faint owing to the spreading of the signal.
On day 280 the ghost components bifurcate with a sharp subcomponent residing on
the trailing edge of the regular MP and IP emission (\eg Fig. 10c which
will be presented later)
followed by a diffuse region of width 1 ms.
We discuss the evolving location, relative amplitude and shape of ghost components 
in \S 6.

\section{NEUTRON STAR ROTATION MODEL}

Table 1 presents the rotation, astrometric and dispersion model parameters
used to align signals for the
images presented in Figures 1 and 3. These parameters lead to reasonably
steady phases for the three months prior to the dispersion event
in late 1997 October. We used template matching software
for arrival time measurements that simultaneously fits for the decay
time of an exponential broadening function which is the result of
multipath propagation. Simultaneous solution for variable
broadening is particularly important for the 327-MHz data. While this
procedure is adequate for 327-MHz, \cite{Rankin73} show how at lower
frequencies one must consider the effects of two scattering screens.
The intrinsic profile that was used as our 
basic template was obtained during an epoch of low scattering.
Our template matching software provided the amplitudes, positions and
scattering times which are plotted in Figures 2 \& 4. 
The arrival time parameters extracted from the data (Fig. 2b \& 4b) 
were used to explore residual timing activity along with dispersion measure
variations. 

A simple extrapolation of the old pulse phases from days 215 to 290
to days 310 to 360 at the two frequencies (Fig. 2, 4)
yields phase jumps in late 1997 October of
5.1 ms and 2.3 ms at 327 MHz and 610 MHz, respectively. If these jumps
were solely the result of dispersion changes, then one could estimate
the change from each frequency record independently.  These estimates
are 0.13 cm$^{-3}$pc and 0.21 cm$^{-3}$pc, respectively. We conclude
that the changes in phase are not simply dispersive.
Note that the effects of scattering have been 
removed in Figures 2b and 4b, and therefore we can't appeal to
scattering as the source of the discrepancy. In fact, scattering
would produce the opposite sense of discrepancy in the inferred dispersion measure 
jumps; \ie we would infer a larger DM jump for 327 MHz relative to 
610 MHz. We conclude that there was an intrinsic slowing 
down\footnote{pulsar arrival times and model residuals are defined such that
an increase in timing residual, or phase, relative to the model corresponds 
to a slowing down of the parent star.} of the rotational phase of the
pulsar around the time of the dispersion jump. This conclusion
is in fact supported by independent measurements of the Crab pulsar at 
1.4 GHz \cite[]{Smith00}. The phase jumps
given above yield a slow down of 1.2 ms along with a dispersion increase
of 0.10 cm$^{-3}$pc. This leaves us with the unsatisfying coincidence of
an unusual intrinsic timing event which is necessarily internal to the 
neutron star with a bizarre propagation event which is most naturally
ascribed to plasma structures 1-2 pc distant from the pulsar.

The deceleration of the pulsar seems to have started around day 290 as we can 
just follow the old and lightly scattered pulse emission while the
amplitude is extinguished during the onset of the
dispersion event (Fig. 1,3). Fits to the interpulse position at both 
frequencies during this critical period are shown in Figure 5a.
By focusing on just the data near the old interpulse position we were
able to follow its location better than our general purpose template
matching software.
The 610-MHz arrival times have been adjusted by a constant
45 bins, or 1.2 ms, in this display. This shift is nominally the
consequence of a dispersion measure difference from the assumed
model (Table 1) of 0.14 cm$^{-3}$pc. Evident in Figure 5 is a frequency
independent slow down of 0.22 ms during days 290-300 that
we conclude is what begins to establish the
new phase of the pulsar as the new emission appears at the
new dispersion measure. Timing noise of the 
Crab pulsar is characterized by a random walk in frequency with
an amplitude that leads to 0.3 ms changes on time scales of a month;
\eg see days 220-280 in Figure 5. The occurrence of an unusual
spindown event -- large amplitude in short time -- at the very moment 
of the dispersion jump and
amplitude null requires serious consideration about the possible
coupling of these phenomena. 

On MJD 50812 (1997 December 30) a ``conventional'' timing
glitch occurred with $\delta\nu/\nu=9\times 10^{-9}$
\cite[]{Wong00}. The decay
timescale of the transient part of the glitch, 2.8~d, is also typical of 
past glitches.  The beginning of this glitch is just detectable in the MP/IP
on the last day of data presented in Figure 3.

In conclusion there are two timing events attributable to internal activity
in the neutron star during this otherwise plasma propagation event. 
The probability of this ``coincidence'' is difficult to assess. 
Wong \etal report that the Crab has been involved in a cluster of rotation
events during 1995-1999.  This heightened ``internal'' activity coincides
with the extended episode of large dispersion and scattering activity
during which the particular events discussed in this paper have occurred.
We proceed in this paper with the assumption that the coincidence of the
two timing events with the dispersion/scattering events 
is the result of chance. While this
assumption is plausible given the excess recent activities in both
properties of the Crab emission, in a future analysis
one might choose to explore a causal link between these seemingly disparate
phenomena.

\section{PLASMA PRISM MODEL}
\subsection{Basic model for DM changes}

After removal of the neutron star rotation model which was presented in
the previous section from the timing data, we fit the
residuals for time variation of the dispersion measure (DM). The results
are shown in Figure 6. The DM variations are characterized by a 
jump of $\delta$DM$_\circ\simeq$0.12 cm$^{-3}$pc that coincides with
the interval around MJD 50700 when pulses are present at both dispersions
as discussed in \S 3. Around the time of the jump the results presented
in Figure 5 provide more accuracy owing to the careful fitting of individual
pulse components.
After the jump the DM steadily declines with most of the jump amplitude lost
over $T\sim 250$ days. There are small increases centered on MJDs 50825 
and 50890 (Fig. 6).  A slower decline has continued to MJD 51300.
We interpret the sudden rise and steady decline as evidence for the
passage of a uniform density triangular prism of plasma through our sight line 
to the pulsar. One lateral face of the prism must be parallel to the line
of sight to provide the observed jump when the geometric sight line
enters the prism owing to relative motions of pulsar, prism and observer
(see inset in Fig. 7). The physical nature of the prism will be addressed
briefly in our conclusion.
Here we are concerned with the radio wave propagation through
the prism as it crosses our sight line to the pulsar.

The steady DM gradient during the 250-day interval after the DM jump
can be converted into a frequency dependent
refraction angle $\theta_r(\nu)$ in the prism. 
The triangular plasma prism model has three important
parameters: the extent transverse to the sight line $L$ which
we equate to the product of the motion of the
sight line transverse to the pulsar direction $V_\perp$ and $T$;
the extent along the sight line at the ``thick'' end of the prism $fL$, and an 
excess electron density $\delta n_e\equiv\delta$DM$_\circ/fL$. The
two length measurements are the height and the base of the assumed
isosceles triangular cross section of the prism. The effect
of this prism on propagation can be accurately calculated based on the
equivalent phase screen $\Phi(x,y)$ that is derived by line of sight
integration of the electrical phase through the dispersive prism:
$$\Phi(t)\equiv\lambda r_e~\delta {\rm DM(t)},\eqno(1)$$
which leads to the refraction angle
$$\theta_r=k^{-1}\nabla\Phi \eqno(2a)$$
$$\theta_r={\lambda^2r_e~\delta{\rm DM}_\circ\over 2\pi V_\perp T}. \eqno(2b)$$
Quantitatively the refraction angle through the prism at 327 MHz is
$$\theta_r(327{\rm ~MHz})=0.35~\mu{\rm rad~} V_{\perp,100}^{-1}, \eqno(3)$$
where the perpendicular velocity is expressed in units of 100 km s$^{-1}$.
The phase velocity in cold plasma exceeds c and therefore the sign of the
refraction is such that an {\it overlap} in time is expected for the direct 
and refracted signals, which is what we reported in the previous section.
We return to this point in the following section that presents a simulation
of the optics.

The geometric time delay, $\delta t_r$, resulting from this refraction 
is very likely small in comparison with the several-ms dispersion delay (Fig. 5):
$$\delta t_r^{327}=z(1-z)D\theta_r^2/c=20~\mu{\rm s~}z_{0.001}V_{\perp,100}^{-2}, \eqno(4)$$
where the location of the prism as a fraction of the distance between the pulsar
and the Earth $z$ is expressed in units of 0.001 (2 pc), appropriate
for the filamentary interface around the Crab synchrotron nebula.
The frequency dependence of this refractive delay is nominally $\nu^{-4}$,
and therefore in future events it could be separated from a dispersion delay 
with sufficient frequency sampling. 

Refraction in the prism will lead to a signal path that depends on radio 
frequency.
The transverse displacement of the image path at the location of the perturbing
plasma is $zD\theta_r\sim 2\times 10^{12}{\rm ~cm~} z_{0.001} V_{\perp,100}^{-1}$
at 327 MHz. This displacement toward the thin end of the prism
results in a smaller dispersion measure for 327 MHz
relative to that for 610 MHz at any instant. An estimate of the size
of this effect using this displacement and
the observed gradient is 
0.0011 cm$^{-3}$pc $z_{0.001} V_{\perp,100}^{-2}$ which also contributes a
small time delay of 40 $\mu$s $V_{\perp,100}^{-2}z_{0.001}$. 
This effect will not contribute significantly to the 
non-dispersive time delays estimated from the extrapolated rotation model of 
the neutron star presented
in the previous section if the site of the dispersion changes is in the
filament zone of the nebula.

For an aspect ratio $f=1$ and $V_\perp=100$ km s$^{-1}$ the excess electron 
density in
the prism is 1700 cm$^{-3}$, and the emission measure for this choice 
of parameters is small, $\sim$2 cm$^{-6}$ pc, relative to that of the
filaments. If the prism extent along the sight line is several times that
in the transverse direction ($f\sim 3$) and if the transverse velocity is as large
as 300 km s$^{-1}$, then the prism density can be as low as 170 cm$^{-3}$.
As the densities of the ionized filaments which have scale size
of $10^{16-17}$ cm are estimated to be around 1000 cm$^{-3}$,
a much smaller density perturbation on the much smaller scale of the
prism is likely, and possible, given this discussion of geometry and motion.
Thus a factor of $\sim 3$ is favored.

\subsection{Transverse velocity from refractive amplitude variations}

The variable amplitude of the pulsed emission
can be used to constrain the transverse motion
of the plasma prism if we attribute these changes to refractive focusing
and defocusing. 
\cite{Clegg98} provide a useful description of the refractive optics
of a Gaussian plasma lens. The refractive gain is assumed to be
single valued (near field) and is a function of the
second derivative of the phase accumulated after propagation through
the lens. The amplitude drops into a deep null over about 10 days 
(290 to 300) at 327 MHz 
and over a shorter interval at 610 MHz (Fig. 2c,4c). We
attribute this to the spreading of rays by the defocusing power of
the phase screen just ahead of the plasma prism.
Following \cite{Clegg98}
the gain which results from a refractive 1D parabolic phase variation located 
at $zD$ along the sight line from the pulsar is
$$G_x=[1 - z(1-z)Dk^{-1}\nabla^2\Phi]^{-1}. \eqno(5)$$
With $\nabla^2\Phi$ estimated from 
$\delta{\rm DM}_\circ/(V_\perp~\delta T)^2$ and a gain of 0.1-0.2 
and $\delta T=10$ d, we estimate  that the transverse velocity is
$$\hat V_\perp = 150 {\rm~km~s}^{-1}~\sqrt{z_{0.001}}. \eqno(6)$$
The shorter time scale for the decrease at 610 MHz is consistent with
expectations of this simple model. Furthermore, there is evidence for
a parabolic term in the phase owing to the observed variations of dispersion
measure in Figures 5 \& 6 along with its conversion to phase (eq. 1).
For our model prism the dominant parabolic phase variation is most
likely directed normal to the projected edge of the prism and
not along the direction of the motion of the sight line; see Figure 7.
The velocity estimated from this calculation is then a lower limit to
the true transverse motion of the prism. 

What is the transverse motion of the sight line to the pulsar relative
to the possible filamentary material in front of it? If the material
is just radially expanding, then there would be one contribution
to the transverse motion from
the pulsar motion, 135 km s$^{-1}$ \cite[]{Wycoff77,Caraveo99}.
This will be reduced by a component of the radial motion of the filaments
which results from the misalignment
of the current pulsar location from the expansion center. For the 
estimated 10$^{\prime\prime}$ misalignment \cite[]{Trimble68}
this amounts to 100 km s$^{-1}$ for material at 1.5 pc moving
at 1500 km s$^{-1}$. The net motion is then only 35 km s$^{-1}$.
There will be an additional contribution from the random velocity of 
the filaments causing the propagation effects discussed in this
paper, assuming that they are indeed the site of the perturbations.
\cite{Trimble68} makes a strong case that the non-radial motions
of the isolated filaments which she studied
are less than 300 km s$^{-1}$ and are typically 70 km s$^{-1}$. 
Her reasoning is that a well defined convergent point
of expansion would not be found if non-radial motions were larger.
The further conclusion from her study is that the expansion velocities
range from about 500 km s$^{-1}$ up to 1500 km s$^{-1}$ which corresponds
to radii from 0.5 pc to 1.5 pc, respectively. The filaments
nearest the pulsar in Trimble's study -- numbered 158, 160, 208 and 209 -- have
observed transverse motions of 47 to 267 km s$^{-1}$. 

More recent studies of the Crab nebula filamentary material have come
to the conclusion that many features can be attributed to Rayleigh-Taylor
MHD instabilities \cite[]{Chevalier75,Hester96,Sankrit98}. The low-density,
synchrotron emitting matter driven by the pulsar is pressing on the
denser supernova ejecta. Fingers or sheets of ejecta matter drip
into the interior along quasi-radial paths. These authors estimate
velocites of the shock interface of 150 km s$^{-1}$. The Alfv\`en
velocity is significantly smaller, $v_{\rm A}=$20 km s$^{-1}B_{-3.5}\sqrt{Zn_3}$,
where the magnetic field $B_{-4.5}$ is in units of $3\times 10^{-4}$ G, the density
$n{_3}$ is in units of $10^3$ cm$^{-3}$ and the atomic number $Z$
may be as large as 4 
if the relevant filament is dominated by Helium \cite[]{Uomoto87}.
Both the shock velocity
and a component of the velocity from any non-radial motion
of the unstable interface could contribute to the transverse velocity
that we need to convert our temporal record to a spatial record.
A second epoch of HST imaging
is planned that will allow much higher resolution study of motions of
the small scale filamentary features (Hester, 1999 personal communication). 

On a larger scale most of the filamentary material which would be along our
sight line to the pulsar has been associated with a constricted
toroidal region around the synchrotron nebula that is associated with
the `dark bays' in the synchrotron emission \cite[]{Fesen92, Lawrence95}.
These authors suggest the possibility that this is the result of
a circumstellar disk that predated the Crab SN. While larger transverse
velocities might be associated with this morphology, Trimble's observational
limits on motions of specific filamentary features remains.
In summary, we choose to use 200 km s$^{-1}$ for our length to time conversions.

The amplitude of the `old' signal has a broad maximum 
that extends over 40 (20) days centered on day 270 (285)
at 327 (610) MHz, respectively (see Fig. 2c,4c). 
There is some evidence for two local maxima
with dips at day 270 (281) and preceding and following higher values.
This peaking, and possibly double peaking, of the amplitude
has the appearance of a caustic crossing event that can be associated
with the subsequent sharp loss of amplitude around the time of the dispersion
jump which was discussed earlier in our estimate of the transverse
velocity. \cite{Clegg98} analyze the optics of a Gaussian plasma lens and
show how it can reproduce the amplitude signature of extreme scattering
events. The equivalent lens at the time of the jump which
strongly defocuses the radiation will lead to a pileup of amplitude
prior to the jump. The absence of a similar caustic crossing event on the
egress can be attributed to asymmetries of the equivalent phase screen
in the two transverse dimensions. 

\subsection{Broadening from multipath propagation}

The new pulse viewed through the prism is broadened at 327 MHz by an exponential function
with a typical 
time constant of $\tau_s=2.1$ ms (\S 3; Fig. 2c). If we equate this broadening to the
simple thin screen result
$$\tau_s = z(1-z)D\theta_s^2/c \eqno(7)$$
with $\theta_s=\lambda/l_\circ$ as the scattering angle of the screen, then
the coherence length scale of the screen is $l_\circ=3\times 10^3$ cm for 327 
MHz and $z=0.001$. At $z=0.001$ the scale of the first Fresnel zone is 
$a_{\rm F}=2\times 10^{10}$ cm,
and, owing to the inequality $l_\circ << a_{\rm F}$, we expect strong diffraction effects.
The same condition holds at 610 MHz.
Phase fluctuations on this small scale can result from either a 
turbulence spectrum extending down to this scale or from gradients of structures
on larger scales. 

The scaling of the new pulse broadening can indicate the slope of the
`turbulence' (average density fluctuation) spectrum \cite{Rickett88}.
Using the results from Figures 2c \& 4c we find an electron density
power law slope of -6, much steeper than the Kolmogorov slope of 11/3
and steeper than the slope expected if we are sampling beyond the
wavenumber cutoff or inner scale which is -4. An alternative interpretation
is that the sight lines are sampling different material (Fig. 7). However
the ratio of pulse broadening at the two frequencies does not change
significantly during 1997 November and December. If there were
significant variations transverse to the path of line of sight through
the prism, one would expect comparable variations along the line of sight.
A further alternative is the idea raised by Cordes
(1998, personal communication) that the fluctuations are intermittent, and
therefore the full extent of broadening is not equally
attained at the two frequencies. The signature of this effect would be
a shallower dependence of pulse broadening on frequency than $-4$. But this model 
would also predict truncated exponential pulse broadening functions and large
variability which are not observed. In future observations during a similar
episode of enhanced scattering multi-wavelength studies are essential
to establish the frequency dependence of pulse broadening.

\cite{Hester96} show that at high angular resolution
many filaments appear to be smooth Rayleigh-Taylor fingers 
which are falling into the synchrotron nebula. In their model these fingers
are stabilized by a transverse magnetic field in the interface
against breakup into smaller scales by Kelvin-Helmholtz instability. 
On the other hand, they also show some filaments 
with fine structure indicative of Kelvin-Helmholtz instabilities. 
The steep fluctuation spectrum in the prism inferred above from pulse broadening 
would favor the presence of Kelvin-Helmholtz instabilities continuing to small scales.
\cite{Jun95} provide an important high spatial resolution
simulation of the growth of these MHD instabilities for various geometries and
magnetic field strengths.
In the future we plan to 
compare the small scale column density, scattering and 
Faraday rotation changes of the Crab pulsar with results obtained from these computer
simulations.  

\section{MODEL FOR THE GHOST EMISSION COMPONENTS}

The ghost emission components described in \S 3 have arrival times that are 
mainly achromatic. This property of the ghost emission 
can be explained if the arrival time is dominated by geometric path delay. 
The plasma prism model described
in \S 5 will lead to multiple imaging, and therefore to multiple pulses if the
resultant geometric relative delays are large with respect to the pulse component width. 
As an aside, if the relative delays were less than the width, interference would
be observed similar to that reported by \cite{Cordes86b} and \cite{Wolszczan87},
and multiple structure might be detected in the narrow giant pulses
\cite[]{Sallmen99}.
We have already described how refraction in a simple plasma prism can explain the presence
of simultaneous emission at new and old dispersion measures simultaneously.
Here we describe how the ghost emission is consistent with it being the result
of a third image which should be present following the `odd number of images'
optics rule that has been frequently applied to gravitational lensing.
The ghost emission components also show evidence for multipath broadening at 
610 MHz,
and yet are not strongly scattered at 327 MHz. The model discussed below
proposes a possible solution to this discrepancy.

Our explanation for the ghost emission requires further
analysis of the phase screen that results from the plasma prism
model introduced  in \S 5. The dispersion record and the amplitude
modulation indicated the presence of linear and quadratic properties
of the screen both ahead of and in the prism. The full phase screen
required for analysis of wave propagation includes effects of both
propagation through the plasma and the geometric path length. 
The geometric path length contributes a parabolic ``Fresnel
bowl'' of phase centered on the sight line:
$$\Phi_{\rm F}(x,y)=A[(x-V_\perp t)^2+ y^2], \eqno(8)$$
where $A\equiv \pi/z(1-z)\lambda D$ establishes the scale of
the quadratic Fresnel phase bowl in radians by dividing by the square of the 
radius of the first Fresnel zone. 
We place the leading (thick) edge of plasma prism at $(x=b,y=0)$ with an 
orientation angle in the $xy$-plane of $\zeta$ with respect to the $x$-axis. 
The equation for the prism leading edge is then $y_p=\tan\zeta\cdot(x-b)$.
We give the prism leading edge a sinusoidal density
profile which leads to sinusoidal phase profile:
$$\Phi_{\rm P}={B\over 2}\left (1-\sin[\pi{(x-b)\over w}\cos\zeta+
\pi({y\over w})\sin\zeta]\right ), \eqno(9)$$
where the half width of the prism's leading edge $w$ is 
$V_\perp~\delta T$ which defines the edge width $\delta T$, 
and $B\equiv\lambda r_e~\delta {\rm DM}_\circ$ 
sets the phase scale of the prism (see eq. 1).
A cross cut of the total phase relative to the geometric line of sight
at $x-V_\perp t$ is shown in Figure 8. If we set $y=\zeta=0$, then
$$\nabla\Phi=2a(x-V_\perp t)-{\pi b\over 2 w}\cos\left [\pi({x-b\over w})\right ]=0 
\eqno(10)$$
establishes points of stationary
phase as a function of frequency. These stationary phase
points can then be used to calculate the corresponding geometric delays 
as a function of frequency and time. 
The roots of $x-V_\perp t$ on either side of $b$ which give
the locations of stationary phases $(x-V_\perp t)_s$ are labeled
with II,III in Figure 8. Signals from point II pass through the leading edge 
of the prism, and therefore have a dispersive delay 
similar to that of the old emission; point III signals pass through the prism 
and have the extra dispersion of the prism.
Nominally these points are close to $x-V_\perp t=b$ where the geometric
delay is
$$t_{\rm ghost,geo}=z(1-z)(x-V_\perp t)_s^2/cD. \eqno(11)$$
Association of image II with the ghost emission
explains the quadratic and nearly achromatic properties of what has been 
observed.

This model of the optics was used to simulate a sequence of pulse
arrival times. The dispersion record was used
to estimate $\Phi_{\rm P}$ with the following elements: a constant;
followed by a cosinusoidal decrease over 60 days preceding the jump
with an amplitude of 0.01 cm$^{-3}$pc; followed by a cosinusoidal increase 
over 15 days with an amplitude of 0.1 cm$^{-3}$pc; and a final linear decay
over 250 days. 
This phase screen was added to the quadratic Fresnel term with the screen placed
at $z=0.001$. The stationary phase points were then located for each day
using a transverse velocity of 200 km s$^{-1}$. At each stationary
phase point dispersive delays at the two frequencies
and the geometric delay were tabulated. 
Figure 9 displays the relative arrival times of the observed
pulses at the two frequencies as a function of time
using the enumeration of stationary phase points in Figure 8. 
The simulation, 
which was obtained with very little iteration of the model parameters,
is reasonably consistent with the observations in Figures 1-4.
Note that the 327-MHz ghost emission arrives slightly earlier
than that at 610 MHz owing in part to the 
decrease in dispersion measure prior to the moment when the
geometric sight line enters the prism.

The absence of the pulse broadening
effects of scattering (diffraction) in the 327-MHz ghost emission
is curious given the 
scattering which is evident in the 610-MHz ghost emission (Fig. 4c). 
The plasma prism model analyzed
here only includes linear and quadratic phase
terms. The prism itself is very turbulent as shown by the pulse
broadening of the new signal (Fig. 2c,4c). The stationary phase points for the 
ghost at 327 MHz (II) will pass through less of the prism than at
610 MHz which can contribute to the absence. 
Furthermore it may be difficult to detect parts of the signal that are very
heavily
scattered at 327 MHz. The very weak amplitude of the 327-MHz ghost emission
relative to 610 MHz (\S 3) suggests that we are observing 
along a low gain path which has the correct refraction angle for us to 
observe and has negligible diffractive scattering.

Figure 10 provides three examples of 610-MHz ghost components. 
The regular MP and IP emission has been subtracted using a template of the
pulse derived from earlier observations (Table 2) along with the
exponential broadening from scattering as determined at 327 MHz
(Fig. 2c). The scales are set such that the ``old'' MP component
peak amplitude is 1.0. Subtraction of the old emission is not always
perfect at the level of 5\%. Owing to the narrow width of the MP
and IP, we are confident that the imperfect subtraction
has no effect on the ghost emission
profile in these data. The result for day 270 demonstrates a typical
result: the ghost of the MP is identical in shape, relative position
and relative amplitude to that of the IP. This provides strong
support for the conclusion that the ghost emission is a replica of
the entire pulse delayed by extra path length along which the
scattering properties are significantly different to that of the
direct path.

However there are days when the MP and IP ghost emissions are dissimilar.
On day 262, the ratio of the IP ghost emission to that of the main
pulse is much less than expected from the delayed and scattered replica
hypothesis. On day 293 the shapes of the MP and IP ghost emission are
dissimilar -- the IP extends longer than the MP. The shape of either
of the ghost components is not simply an exponentially broadened
version of the undelayed counterpart. Within the context of our
plasma prism model non-exponential scattering profiles are simple
to explain. For example, Cordes (1998, personal communication) has
explored the effects of small-scale intermittency on scattered pulses.
By small scale he means that the turbulent plasma does not fill
the diffractive scale, $l_d=z\theta_s$. If so, one or more highly
truncated exponential pulses would be observed from individual
``screenlets'' with their associated geometric delays and scattering
time constants. This incomplete scattering situation was also discussed 
by \cite{Lyne75} to explain the odd pulse shapes observed
during the earlier episode of extreme scattering.

The occasionally dissimilar shape of the MP and IP ghost components may 
indicate that the phase screen is effectively resolving their angular separation. 
MP and IP emission is most likely from the outer magnetosphere. \cite{Romani95}
and \cite{Harding98} provide recent models of the high frequency
emission, and one generally assumes that the sites of the MP and IP radio emission 
are coincident with the sites of the corresponding high frequency components. 
In the \citeauthor{Romani95} model
the MP and IP will have a projected separation at the plane of
the neutron star center which is comparable to that of the light
cylinder radius, 1500 km. At the distance of our hypothesized phase
screen, 1-1.5 pc, this corresponds to 7 $\mu$as. In the \citeauthor{Harding98}
polar cap model no such offset is expected. In \S 5 we estimated
that the average refraction angle in the plasma prism is of order
0.35 $\mu$as at 327 MHz (eq. 3) which is then $\sim 0.1~\mu$as at 610 MHz. The 
presence of significantly larger gradients in smaller regions might be possible. 
If so, these could lead to significantly different refractive gains for 
the ghost MP and IP emission. Note that while the separation between paths
from the MP and IP may be much larger than $l_\circ\sim 10^{3-4}$ cm, the size of 
the diffractive disk $z\theta_s(1-z)D\sim 10^{11-12}$ cm is much larger than
the likely path separation. Therefore there is little change expected in
the diffraction of the two pulse components. The perturbed optics of the filamentary 
zone of the nebula during this epoch may thus be resolving the magnetospheric
emission sites and therefore supporting high altitude emission models.

\section{CONCLUSIONS}

In this paper we have:

\noindent
(1) reported on an anomalous events in the record of the Crab pulsar
during the second half of 1997 which involve rapid changes in the
pulsar's flux, dispersion measure and pulse broadening time as well
as the appearance of ghost emission;

\noindent
(2) found what appears to be a slowing down of the rotational phase of
the parent neutron star at the time of the propagation event which we
tentatively conclude is coincidental owing to the current clustering
of rotation events;

\noindent
(3) modeled the flux, dispersion measure and pulse broadening time
changes and the ghost emission as the refractive and diffractive
effects of a plasma prism which crossed the sight line and is located
in the filamentary zone of the Crab nebula; and

\noindent
(4) concluded that the corrupted optics of the plasma prism has provided
sufficient resolution to distinguish the apparent locations of the
main pulse and interpulse which lends support to outer magnetosphere
models.

These results provide important constraints on the small scale
density structures for
detailed simulations of the three dimensional plasma dynamics in 
the filamentary zone of the Crab nebula. The results complement
high resolution studies of line emission which can provide three
dimensional motions and physical conditions of the same material
at reduced spatial resolution.

During the next episode of strongly perturbed propagation a number
of supporting observations are needed. High angular resolution measurements
with VLBI
at wavelengths down to 1m and longer will provide limits on, or measurements
of, both the separation of the direct and ghost emission and 
the apparent angular diameter which will improve the inferences about 
the screen's location
and its transverse motion.  Accurately calibrated polarimetry will
provide measures of Faraday rotation which samples
the radial component of the magnetic field. The nominal overpressure in
the plasma prism could be balanced by magnetic pressure surrounding it
which could lead in turn to a distinct Faraday rotation signature. 
Alternatively the prism may be part of a evolving shock structure.
Our current polarimetry data will be studied for
evidence of variable Faraday rotation, but the offset feed configuration
is known to
have poor polarization properties. Sampling pulse shape and arrival time
over more radio frequencies will allow better assessment of 
dispersive, refractive and diffractive effects. 

\acknowledgments
We are grateful to the NRAO staff for maintenance of the 85ft Pulsar
Monitoring Telescope and for assistance with our observational
program. This monitoring
effort has also been supported through the Naval Research
Laboratory and US Naval Observatory activities with the Green Bank
85ft telescopes. We thank Graham Smith and Andrew Lyne for sharing
the results of their preliminary investigations of the Crab events
using their and our data, and look forward to further synthesis of
all measurements. Our effort has been supported from NSF grants
AST-9307913 in the past and currently AST-9820662.

\bibliography{dbacker}

\setcounter{figure}{0}

\begin{figure}
\plotone{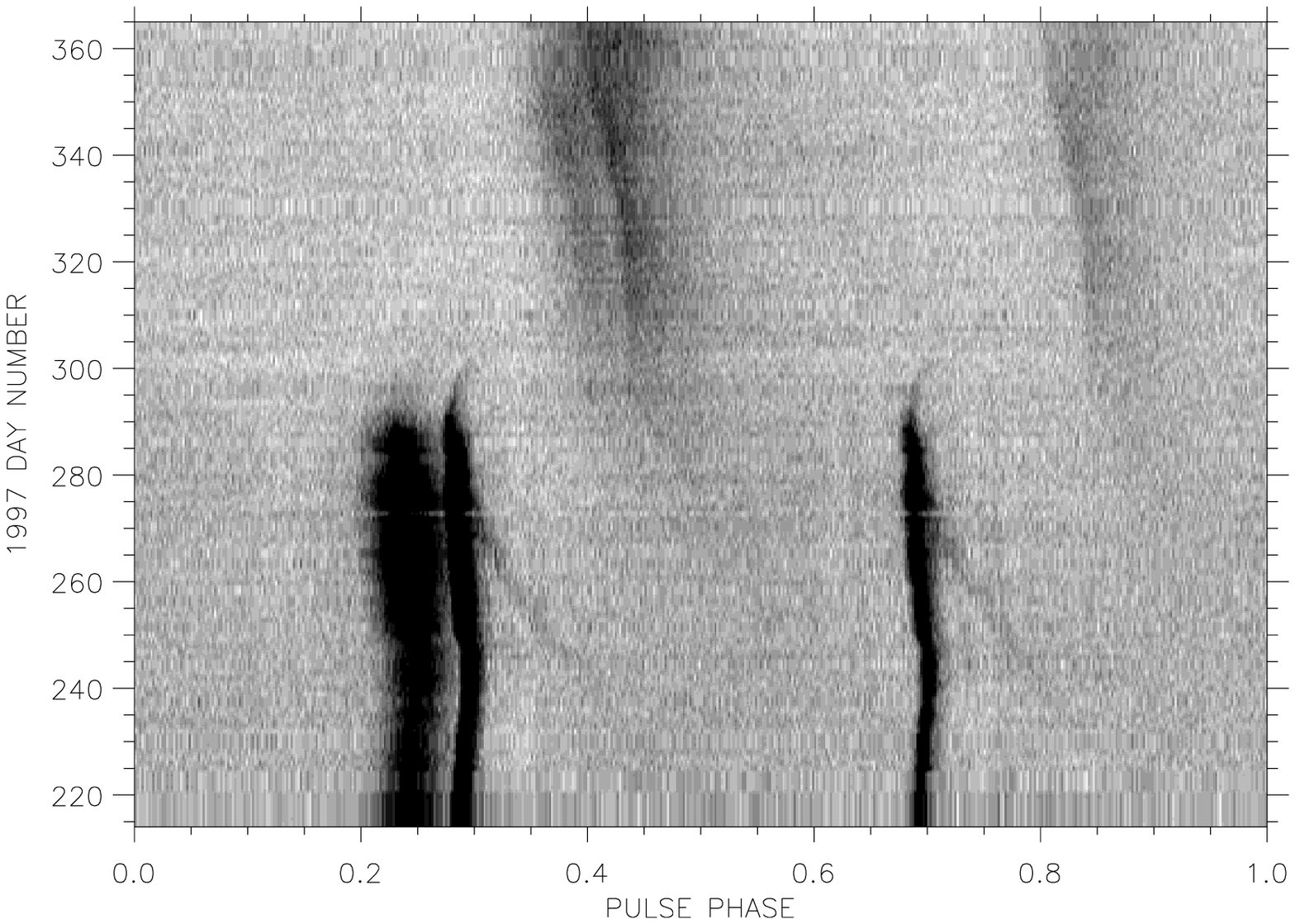}
\caption{Image of pulsar flux at 327 MHz obtained with an 85ft telescope
at NRAO Green Bank. 
The dispersion jump reported
in this paper occurs around day 295, and a rotation glitch has just started on
day 365.
}
\end{figure}

\begin{figure}
\plotone{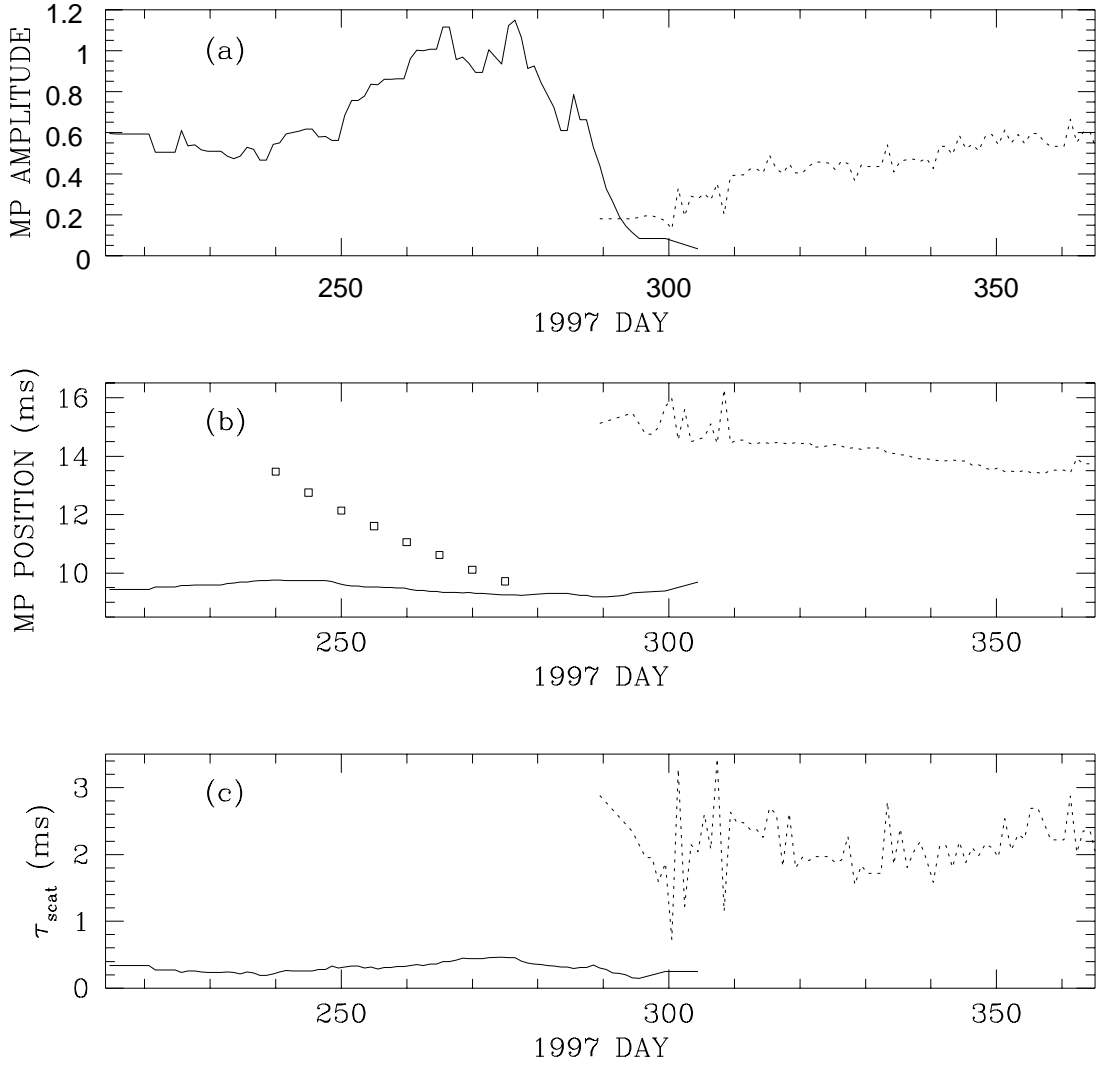}
\caption{Observational parameters of the 327-MHz data. The solid
and dotted lines present parameters of the ``old'' and ``new'' pulses,
respectively: (a) amplitude of main pulse; (b) position of main pulse;
and (c) exponential pulse broadening time scale. In (b) the locations
of the strongest ``ghost'' component emission are shown with squares.
}
\end{figure}

\begin{figure}
\plotone{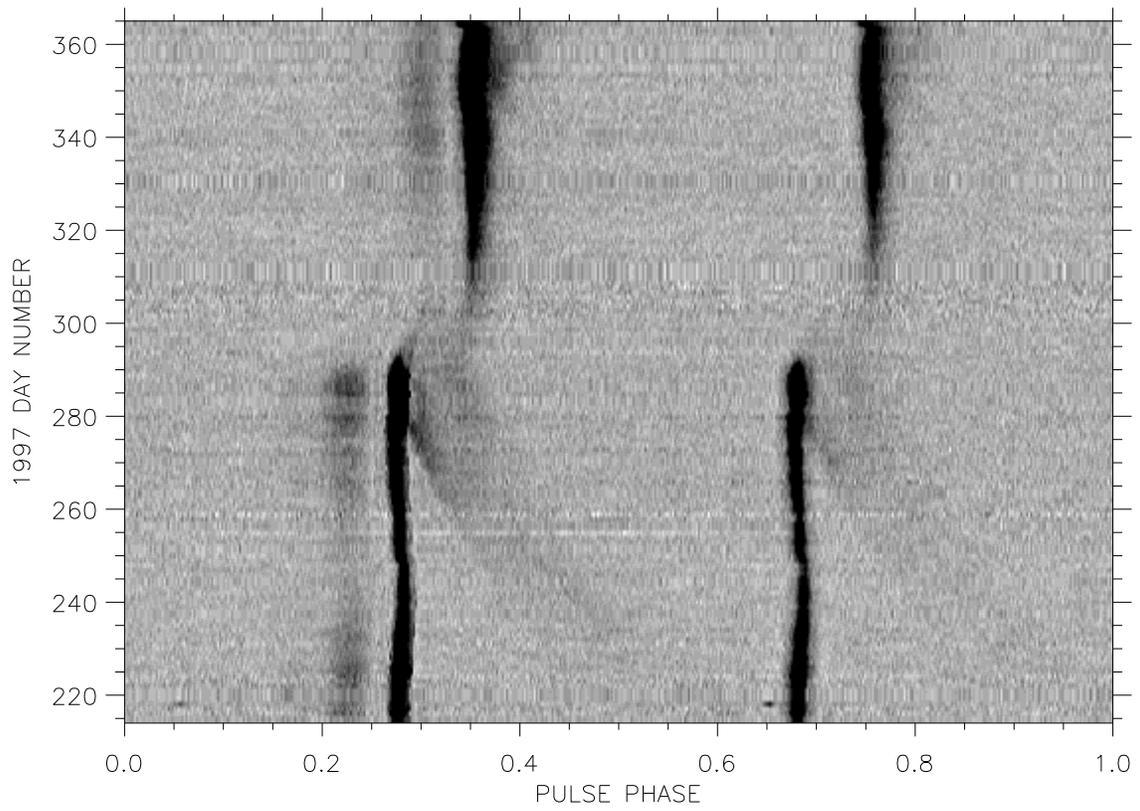}
\caption{Image of pulsar flux at 610 MHz obtained with an 85ft telescope
at NRAO Green Bank. 
The dispersion jump occurs
around day 295, and a rotation glitch has just started on day 365.
}

\end{figure}

\begin{figure}
\plotone{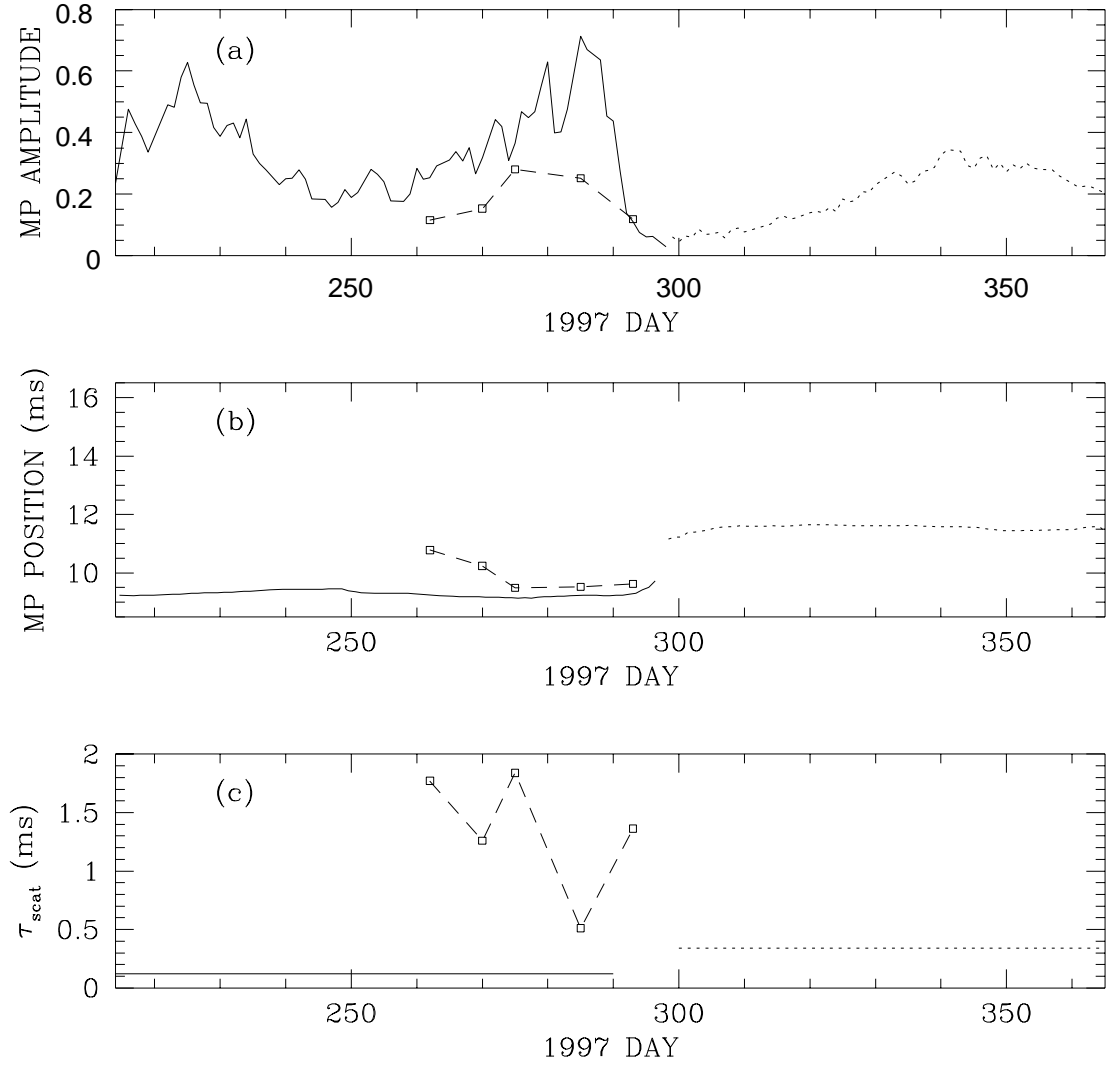}
\caption{Observational parameters of the 610-MHz event. The solid,
dashed and dotted lines present parameters of the ``old'', ``ghost'' and 
``new'' interpulse components, respectively: 
(a) amplitude of main pulse; (b) position of main pulse;
and (c) exponential pulse broadening time scale.
}
\end{figure}

\begin{figure}
\plotone{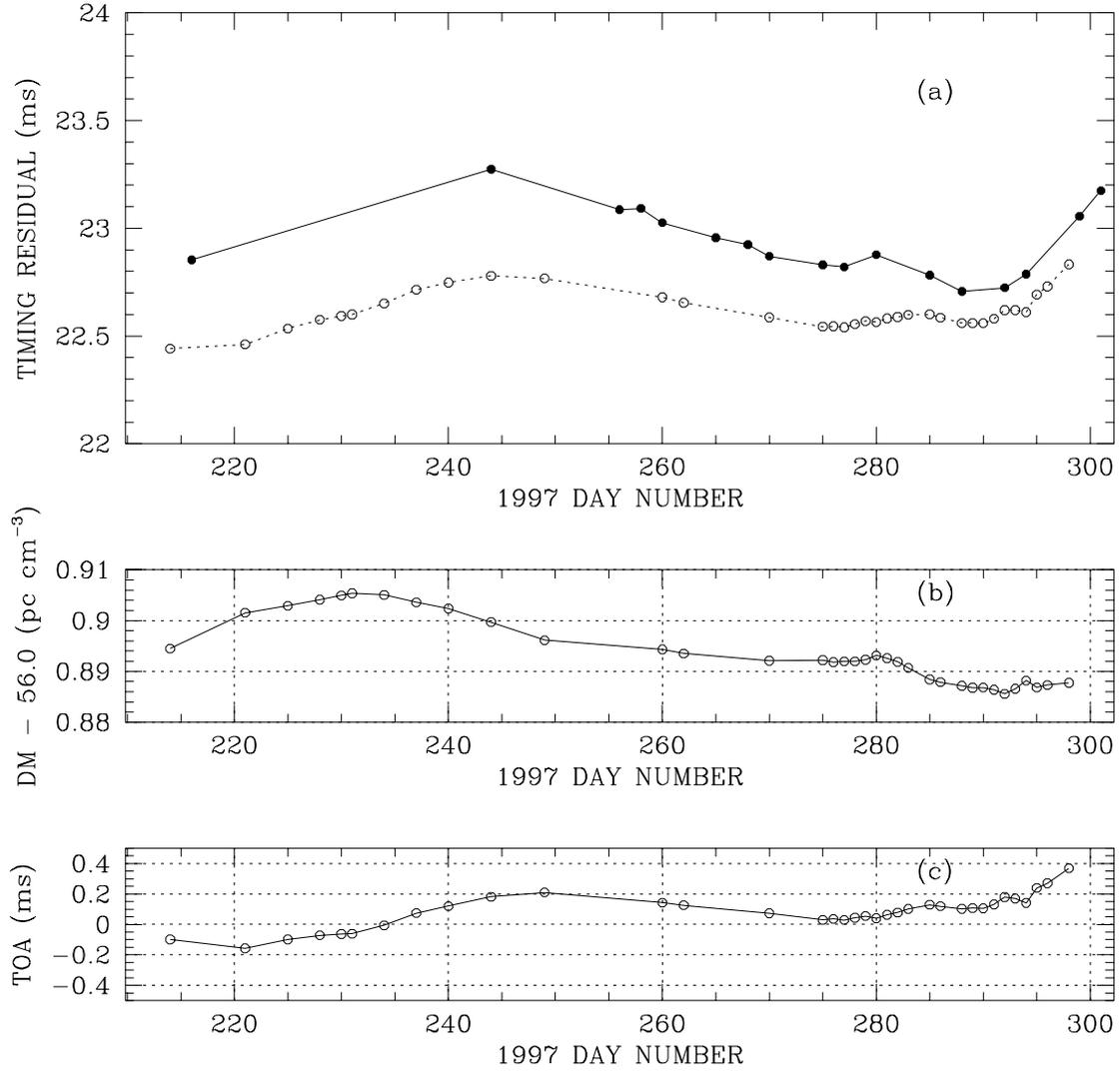}
\caption{In (a) the arrival times at 327 MHz and 610 MHz from Figures
1-4 are compared; the 610 MHz times have been shifted by 1 ms which 
corresponds to a small shift in the dispersion relative to that in
Table 1. The times are then decomposed into residual
dispersion measure in (b) and infinite frequency timing residual
in (c).
}
\end{figure}

\begin{figure}
\plotone{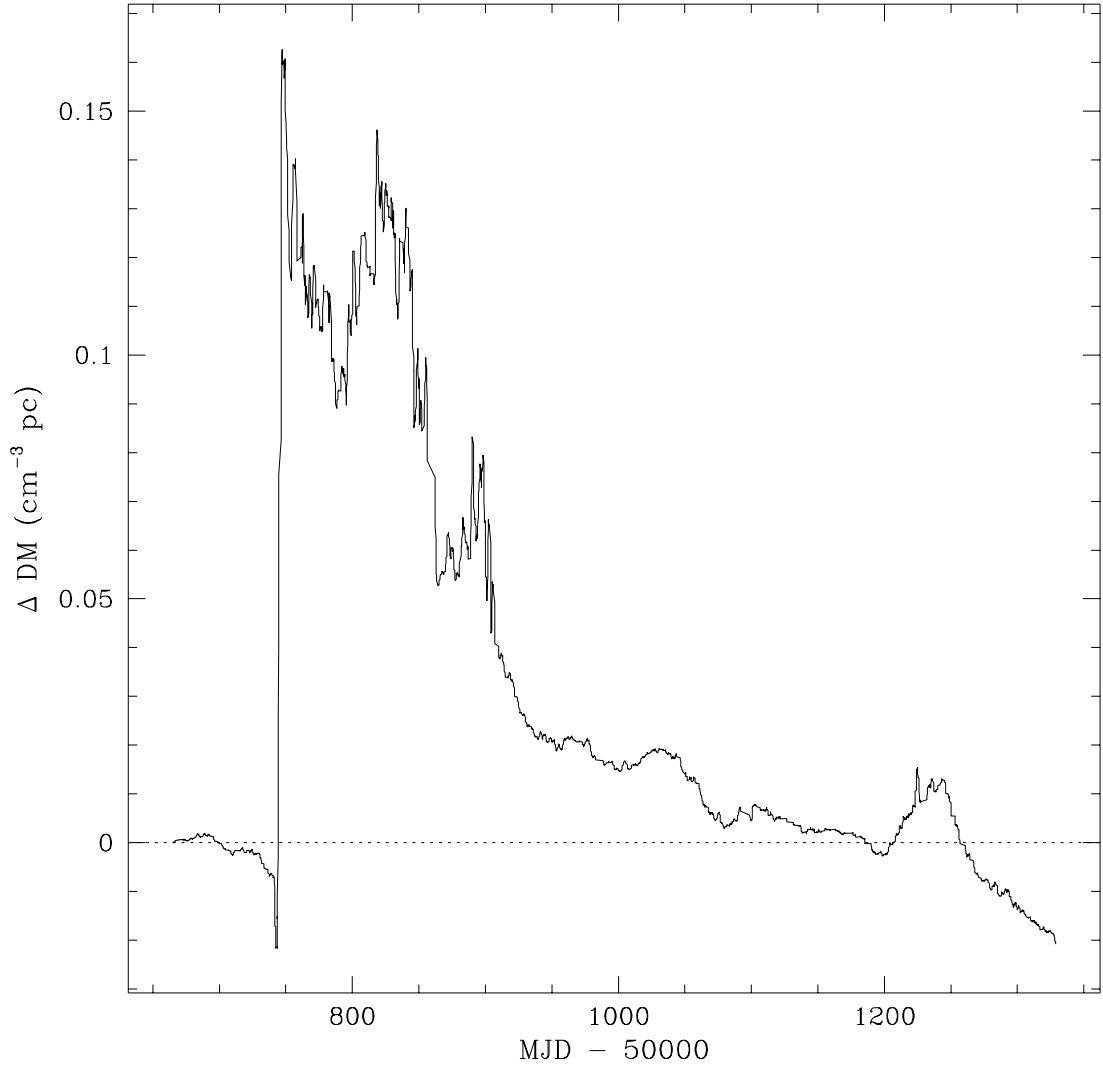}
\caption{Dispersion measure variations from NRAO Green Bank observations
at 327 MHz and 610 MHz. The influence of scattering has been removed. The
data have been smoothed with 7-d running average. The dispersion jump
occurred during MJD 50736-750 (1997 October 15-29; days 288-302). 
Dispersion measurements 
at times closest to the jump are difficult owing to the low flux
and extreme values of exponential pulse broadening (scattering) 
at 327 MHz.
}
\end{figure}

\begin{figure}
\plotone{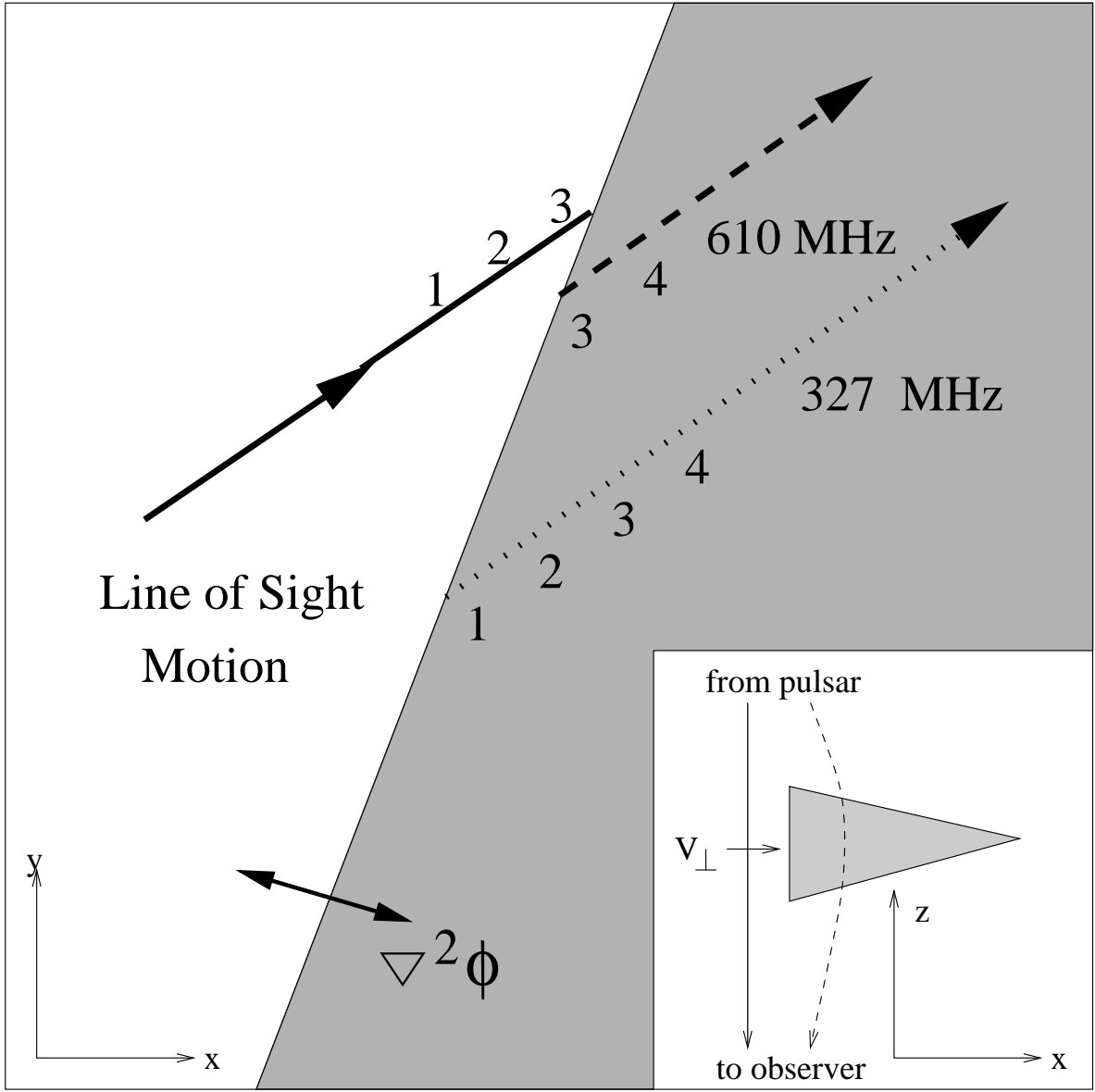}
\caption{Model of motion of sight line and refracted
sight lines at two frequencies across the sky ($xy$ plane)
relative to a plasma prism (shaded). The numbers indicate sequential
locations of the direct and refracted sight lines as they approach
and traverse the prism, respectively. The inset diagram ($xz$ plane)
shows the direct sight line and a sight line refracted through the
triangular prism.
The expected direction of the maximum second derivative of the 
phase screen is shown perpendicular to edge of prism.
}
\end{figure}

\begin{figure}
\plotone{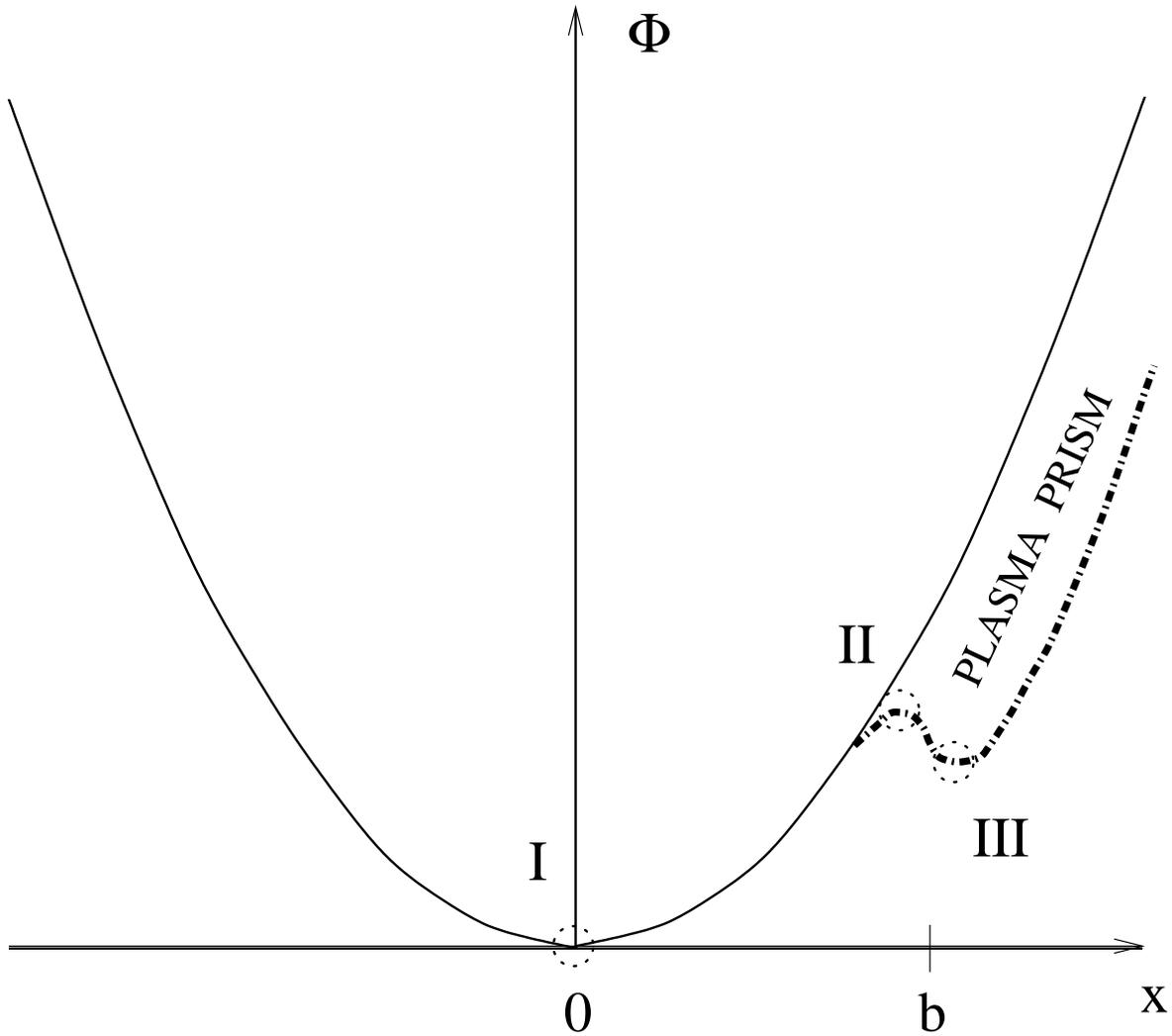}
\caption{Cross section of the phase screen
including a quadratic Fresnel bowl (solid) with the addition of
the leading, sinusoidal edge of prism (dot-dash). Stationary
phase points are interpreted as: I, old emission prior to
DM jump; II, ghost emission; and III, new emission viewed
through the plasma prism. As a function of time the origin
of the quadratic Fresnel bowl moves to the right: toward and then
into the prism.
}
\end{figure}

\begin{figure}
\plotone{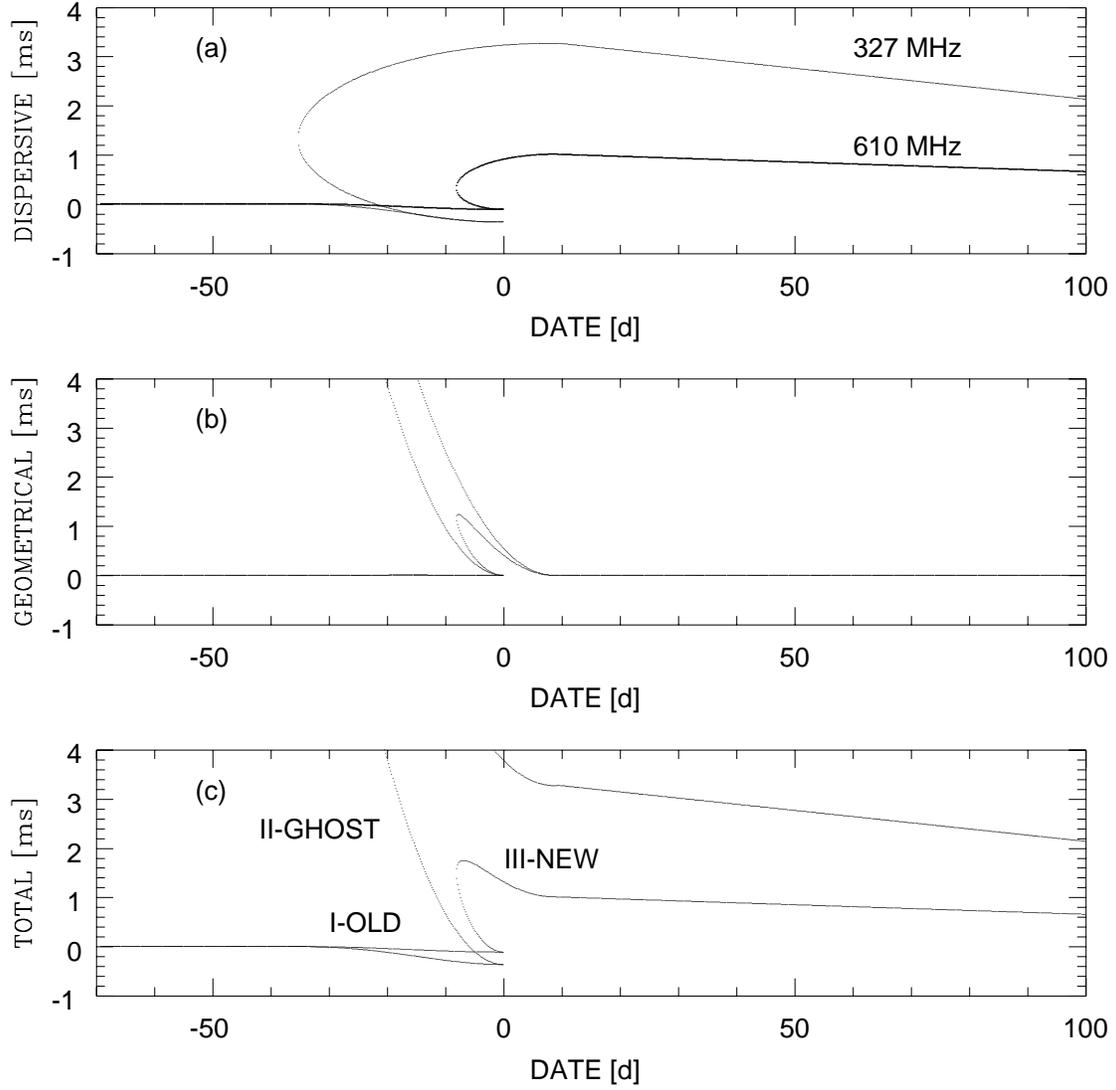}
\caption{Results from a simulation of arrival times for signals passing through 
a plasma prism as discussed in the text. The dispersive component is shown in (a)
the geometrical component in (b) and  the total in (c). The outer
points are at 327 MHz and inner points are at 610 MHz. In (c) the
identification of simultaneous pulse locations with points of
stationary phase in Figure 8 is made.
}
\end{figure}

\begin{figure}
\plotone{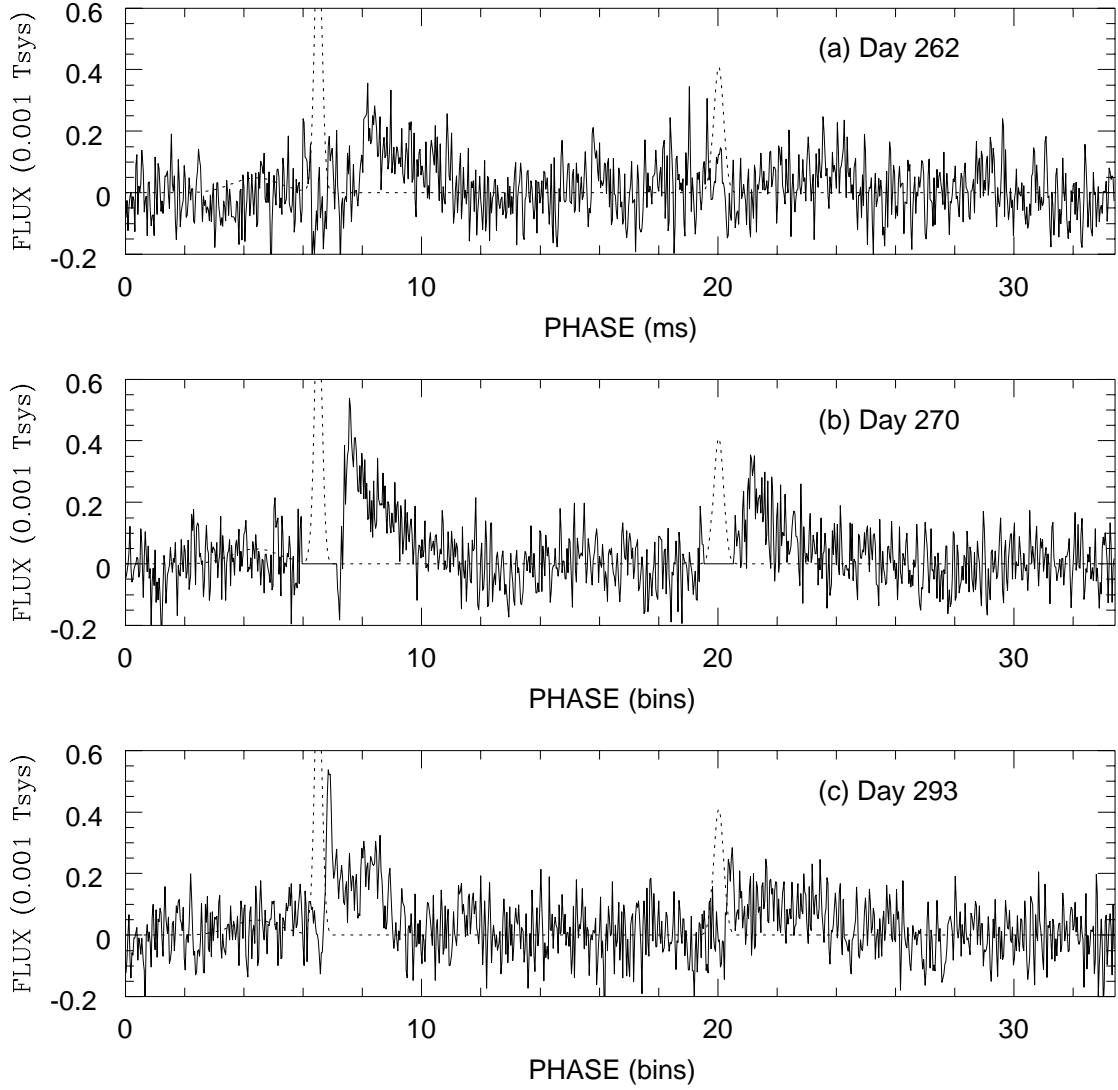}
\caption{Ghost pulse components for days 262, 270 and 293 (solid).
The main emission template is shown with dotted lines. The main
pulse and interpulse components are distorted on day 270 and so
the residuals in these regions are blanked.
}
\end{figure}

\clearpage
\begin{deluxetable}{lc}
\footnotesize
\tablecaption{Crab Pulsar Timing Parameters \label{tbl-1}}
\tablehead{
\colhead{Parameter} & \colhead{Value}   }
\startdata
Right Ascension (J2000){\tablenotemark{a}} & 05 34 31.973       \\
Declination (J2000){\tablenotemark{a}}     & 22 00 52.07	\\
Period (s)              & 0.033458496298951444      \\
Period Derivative (10$^{-15}$s s$^{-1}$) & 420.6590250882 \\
Second Derivative (10$^{-30}$s s$^{-2}$) & -1521098.959 \\
Epoch (MJD)		& 50260.00 \\
Dispersion Measure (cm$^{-3}$pc) & 56.791    \\
\enddata
\tablenotetext{a}{\cite{Lyne93}}
\end{deluxetable}

\begin{deluxetable}{crrrr}
\footnotesize
\tablecaption{Crab Pulsar Profile Model {\tablenotemark{a}} \label{tbl-2}}
\tablehead{
\colhead{Frequency} & \colhead{Component} & \colhead{Relative Phase,$\phi_{\rm n}$} & \colhead{Relative } & \colhead{Width, W$_{\rm n}$}\\ 
\colhead{(MHz)} & 	&  \colhead{(deg)} & \colhead{Amplitude, A$_{\rm n}$} & \colhead{(deg)}
}
\startdata
327{\tablenotemark{b}}	
        & precursor	& 196.5$\pm$ 0.5		& 0.43$\pm$ 0.05 	& 16.4$\pm$ 0.6   \\
	& main pulse	& 214.28$\pm$ 0.05	& 1.50$\pm$ 0.08	& 2.8$\pm$ 0.1 \\
	& interpulse	& 0.0			& 1.00		& 3.5$\pm$ 0.1 \\
	&		&			&		&	     \\
610{\tablenotemark{c}}
	& precursor	& 195.0$\pm$ 0.9		& 0.14$\pm$ 0.02	& 12.0$\pm$ 4.0 \\
	& main pulse	& 214.40$\pm$ 0.08	& 2.08$\pm$ 0.14	& 3.3$\pm$ 0.3 \\
	& interpulse	& 0.0			& 1.00		& 4.1$\pm$ 0.3  \\
\enddata
\tablenotetext{a}{Gaussian component model: $\Sigma$(A$_{\rm n}\sqrt{2\pi{\rm W}_{\rm n}})\exp[-(\phi-\phi_{\rm n})^2/2\sigma_{\rm n}^2]; ~W_{\rm n}\equiv 2\sqrt{2\ln(2)}\sigma_{\rm n}$}
\tablenotetext{b}{From low scattering days: 1995 June 23; 1997 January 12; 1999 May 28; 1999 November 28}
\tablenotetext{c}{From: 1995 June 23; 1997 January 12; 1999 November 28; 1999 December 31}
\end{deluxetable}

\end{document}